\documentclass[12pt,sort&compress]{article}

\usepackage[dvips]{color}
\usepackage{amsmath}
\usepackage{graphicx}
\usepackage{tabularx}
\usepackage{caption}
\usepackage{subcaption}
\usepackage{cite}
\usepackage{amssymb}
\usepackage{mathtools}
\usepackage{bbm}
\usepackage{amsfonts}
\usepackage{tabularx,ragged2e}
\usepackage{xcolor}
\newcolumntype{C}{>{\Centering\arraybackslash}X} 
\usepackage[outercaption]{sidecap}    
\usepackage{hyperref}

\newcommand{\danger}{\mathcal{D}}
\newcommand{\safe}{\mathcal{S}}

\begin{document} 
\title{Winding number of a Brownian particle on a ring under stochastic resetting}
\date{}
\author{Pascal Grange\\
 Xi'an Jiaotong--Liverpool University, Department of Physics\\
111 Ren'ai Rd, 215123 Suzhou, China\\
\normalsize{{\ttfamily{pascal.grange@xjtlu.edu.cn}}}}
\maketitle

\begin{abstract}
We consider a random walker on a ring, subjected to resetting at Poisson-distributed times to the 
  initial position (the walker takes the shortest path along the ring to the initial position at resetting times).
    In the case of a Brownian random walker the 
    mean first-completion time of a turn is expressed in closed form  as a function of the resetting rate.
     The value  is  shorter than in  the ordinary process 
     if the resetting rate is low enough. Moreover, the mean first-completion time of a turn can be minimised in the resetting rate. 
     At large time the distribution of winding numbers does not reach a steady state,
       which is  in contrast with the non-compact case of a Brownian particle under resetting 
    on the real line. The mean total number of turns and the variance of the net number of turns   
    grow linearly with time, with a proportionality constant equal to the inverse of the mean first-completion time of a turn.
\end{abstract}

\pagebreak

\tableofcontents

\section{Introduction}

 In equilibrium statistical mechanics, systems are assumed to have forgotten their initial state. One way  to study the behaviour of 
   a system out of equilibrium is to put it in contact with its initial configuration by resetting it. If the resetting occurs at Poisson-distributed 
   times, the process exhibits a renewal structure. Technically, this renewal structure allows to work out the 
    Laplace transform of the probability of the configurations in the system subjected to resetting, in terms of the 
     probability of the configurations in the ordinary system (we will refer to processes without resetting as {\emph{ordinary}} processes).\\
     
     
      This approach, first taken  in the case of a diffusive random walker on a line \cite{evans2011diffusion}, has led to deep 
       insights on the steady states
      of stochastic processes under resetting, and on  mean first-passage times.
         In particular, the mean first-passage time of a diffusive random walker at a fixed target, 
     which is  known to be infinite in the ordinary case, becomes finite under resetting. Moreover, it can be expressed in closed form 
      as a function of the resetting rate $r$, which exhibits a single minimum \cite{evans2011optimal}. Resetting a Brownian
        random walker to its initial position cuts off 
       long excursions in the wrong direction, which shortens the mean first-passage time on average. On the other hand, resetting may occur  
       when the walker is reset when its position is very close to the target. Intuitively, stochastic resetting decreases the mean first-passage time
        because the amplitude of excursions in the wrong direction is unbounded.
         Stochastic resetting has found applications in a variety of fields, including statistical mechanics and
      active matter\cite{evans2018run,refractory,magoni2020ising,grange2020entropy},  population dynamics \cite{mercado2018lotka,toledo2019predator,ZRPSS,ZRPResetting}, reaction-diffusion systems \cite{durang2014statistical,grange2021aggregation}, search processes
   and stochastic processes \cite{lapeyre2019stochastic,gupta2018stochastic,basu2019long,basu2019symmetric,pal2021inspection,bressloff2020directed,bressloff2020modeling,singh2020random,majumdar2021mean}.
      For a review, see \cite{topical} and references therein.\\
      
        
         In this work we revisit the case of a random walker in one dimension, and put it in a compact setting. 
          The coordinate describing the position of a random walker on a circle is the polar angle, but it comes together  with the winding number of the random walker 
           (the number of turns it has completed since the beginning of the process).
            The distribution of the winding number for a Brownian walker has been studied in \cite{KCM}.
          We first study the mean first-passage time until the completion of a turn (either clockwise or anti-clockwise). 
         The motivation is the topological acceleration effect of resetting: indeed, with two absorbing boundaries at winding numbers $\pm 1$, the
          trajectory of the walker is bounded, and there is no obvious acceleration in the process due to cutting-off  long excursions in a wrong direction. 
           However, resetting the walker to its initial position (going along the shorted path on the circle), may help the random walker 
            complete one turn:  if it is within $\pi$ radians of one of the targets at resetting, it immediately moves to the target.\\

             The paper is organised as follows. In Section 2 we set up notations to describe the acceleration phenomenon. In Section 3 we work out the corresponding renewal equations.
              The Laplace transform of the survival probability (with absorbing boundary conditions at winding number $\pm 1$) 
               is obtained in closed form in terms of the ordinary process. In Section $4$ we work out the quantities needed  
                for the application to a Brownian walker on a ring
                (some of which have already been worked out in \cite{KCM}).
             In Section $5$ we use the explicit expression of the extinction rate of the process under resetting to characterise 
              the long-time behaviour of the winding number of the diffusive random walker under resetting.

\section{Notations and quantities of interest}

Consider a random walker (or particle) on a circle with polar angle $\theta$, starting at $\theta=0$.
  The process ends when the particle completes a turn (clockwise or anti-clockwise).  The duration of the process is therefore the first-passage time
   of the coordinate $\theta$  at $\pm 2\pi$. The system is equivalent to a one-dimensional random walk on a segment with absorbing boundaries at $-2\pi$ and $2\pi$, with the two ends identified by periodic boundary conditions. For an investigation of the mean   time to double absorption under resetting by 
    two  targets on either side of the origin, see \cite{calvert2021searching}.  \\

\begin{figure}[h]
    \centering
    \begin{subfigure}[h]{0.4\textwidth}
        \centering
        \includegraphics[width=2.8in]{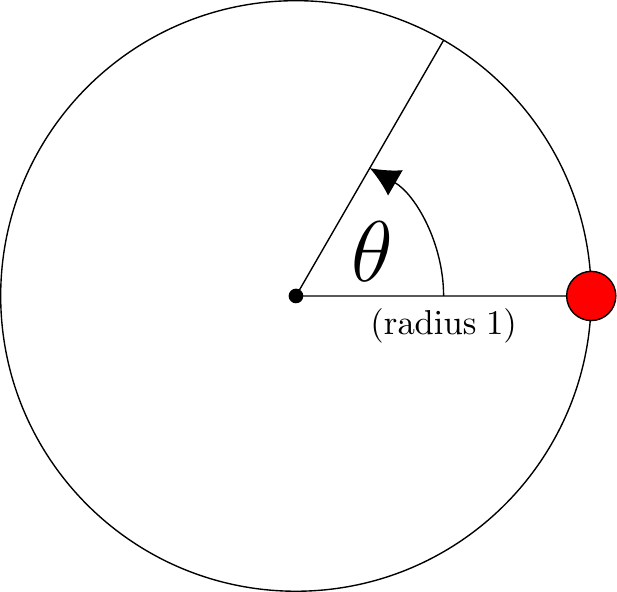}
        \caption{}
    \end{subfigure}
   \hfill
    \begin{subfigure}[h]{0.4\textwidth}
        \centering
        \includegraphics[width=2.8in]{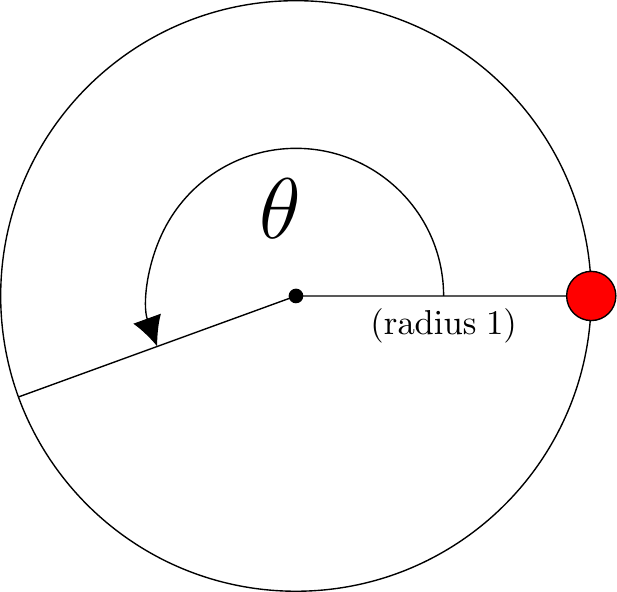}
        \caption{}
    \end{subfigure}
    \caption{Two possible configurations of the system. The process started at $\theta = 0 $. The  patch  symbolises 
     the absorbing boundary conditions at $\theta = \pm 2\pi$. In (a) the particle is in the safe zone, it goes to $\theta=0$ if a resetting event occurs, and the process continues. In (b) the particle is in the dangerous zone: the shortest path to the initial configuration leads to $\theta=2\pi$. If a resetting event occurs in this configuration, the process ends.}
    \label{dangerous}
\end{figure}

When a resetting to the initial position occurs, the particle takes the shortest path
 to its initial position. Hence, if a resetting event occurs when the angle is the domain 
\begin{equation}
\danger:= [  -2\pi , -\pi  ] \cup [ \pi,2\pi  ], 
\end{equation}
 the process ends.
 On the other hand, if a resetting event occurs when the angle is in the domain
 \begin{equation}
\safe:= ] -\pi, \pi  [, 
\end{equation}
 the angle is reset to zero and the walker keeps  going. 
  We will refer to $\danger$ and $\safe$  as the dangerous zone and safe zone respectively. The situation is depicted on Fig \ref{dangerous}.\\

Let us denote by $Q^{(r)}(\theta,t)$ the probability to find the particle at angle $\theta$ at time $t$ in the process
 with resetting rate $r$. The survival probability $q^{(r)}(t)$  of the particle at time $t$ satisfies
 \begin{equation}
  q^{(r)}(t) = \int_{-2\pi}^{2\pi} Q^{(r)}(\theta,t) d\theta.
 \end{equation}
 
  Let us denote by $T^{(r)}$ the duration of the process. Its mean 
     $\langle T^{(r)} \rangle$
   is obtain by integrating the time variable against the 
    absorption rate (or death rate) of the process:
  \begin{equation}\label{qoi}
   \langle T^{(r)} \rangle = \int_0^\infty  t \left(-\frac{dq^{(r)}}{dt}(t)\right) dt= - [ t q^{(r)}( t ) ]_0^\infty + \int_0^\infty q^{(r)}( t ) dt.
   \end{equation}
    As the lifetime of the process is almost surely finite,
   it is enough  to derive the Laplace transform $\tilde{q}^{(r)}$ of the survival probability to express the mean lifetime:
  \begin{equation}
   \langle T^{(r)} \rangle = \tilde{q}^{(r)}(0),
  \end{equation}
 with the following notation for the Laplace transform of any function $f$ of time:
  \begin{equation}
  \tilde{f}(s):= \int_0^\infty f(t) e^{-st}dt.
  \end{equation}

 \section{Renewal equations}\label{renewalSec}

\subsection{Extinction rate of the process}\label{renewalSec}
 
  Let us take advantage of the renewal structure of the process by 
   conditioning on the 
  latest time of resetting in the interval $[0,t]$ (this approach has  proven very efficient in a broad range of models under stochastic resetting, see Section 2.2 of \cite{topical} for a review).\\

  If there is no resetting in $[0,t]$, which is the case with probability $e^{-rt}$, the system evolves 
   according to the ordinary process on $[0,t]$. Otherwise, consider the latest resetting event in this interval, which can be written as $t-\tau$, for some $\tau$ 
   in $[0,t]$. 
    Consider the time-derivative of the survival probability of the process (which is the extinction rate of the process).
   At fixed $\tau$, there is a contribution to the extinction rate  from  histories in which the walker has survived in the process under resetting, undergone resetting at $t-\tau$ while being in the safe zone $\safe$, and evolved according to the ordinary process on $[t-\tau,t]$, to reach one of the targets at time $t$. Moreover, there is a contribution from resetting 
    events at time $t$ that happen when the walker is in the dangerous zone $\danger$ (as in Fig. \ref{dangerous}(b)). These three contributions  add up to
 \begin{equation}\label{evolqr}
 \begin{split}
 \frac{dq^{(r)}}{dt}(t) =& e^{-rt} \frac{dq^{(0)}}{dt}(t)  
  + r \int_0^t d\tau e^{-r\tau} \left( \int_{\safe} Q^{(r)}(\theta,t-\tau) d\theta\right) \frac{dq^{(0)}}{dt}(\tau)
   - r \int_{\danger} Q^{(r)}(\theta,t) d\theta.
  \end{split}      
 \end{equation}
 All the terms in the above equation are negative (as the survival probability decreases with time), hence the negative sign in the last term.\\

   Let   us denote by $S^{(r)}(t)$ the probability of presence of the particle in the safe zone at time $t$:
  \begin{equation}\label{safeDef}
  S^{(r)}(t):= \int_{-\pi}^{\pi} Q^{(r)}(\theta,t) d \theta.
  \end{equation}
 With this notation,
 \begin{equation}
 \begin{split}
 \int_{\danger} Q^{(r)}(\theta, t) d\theta =&  q^{(r)}( t) - S^{(r)}(t)\\
 \end{split}
  \end{equation}
 and Eq. (\ref{evolqr}) reads
  \begin{equation}
  \frac{dq^{(r)}}{dt}(t) = e^{-rt} \frac{dq^{(0)}}{dt}(t)  
  + r \int_0^t d\tau e^{-r\tau}  S^{(r)}(t-\tau)  \frac{dq^{(0)}}{dt}(\tau)
   - r  q^{(r)}( t) + r S^{(r)}(t).
  \end{equation}
 The Laplace transform of the renewal equation for the extinction rate therefore reads
   \begin{equation}\label{renrs}
 \begin{split}
 -1 + s \tilde{q}^{(r)}( s ) =& -1 + (r+s)  \tilde{q}^{(0)}( r+s ) +  r  \tilde{S}^{(r)}( s )\left[ -1 + (r+s)  \tilde{q}^{(0)}( r+ s )\right]\\
    & - r  \tilde{q}^{(r)}( s)   +  r  \tilde{S}^{(r)}(s),
  \end{split}      
 \end{equation}
where we have used the normalisation conditions $q^{(r)}(0) = 1$ and $q^{(0)}(0) = 1$.  In particular,  substituting $0$ to $s$ yields
 \begin{equation}\label{dr0}
 \begin{split}
 0&=  r \tilde{q}^{(0)}( r ) +  r  \tilde{S}^{(r)}( 0 )\left[ -1 + r  \tilde{q}^{(0)}( r )\right]  - r \tilde{q}^{(r)}( 0) + r \tilde{S}^{(r)}(0),\\
 {\mathrm{hence}}\;\;\;\;\;\;\;\;  \tilde{q}^{(r)}( 0)  &=  \tilde{q}^{(0)}( r )\left(  1 + r \tilde{S}^{(r)}(0)  \right).
 \end{split}
 \end{equation}

%
   
   \subsection{Probability density of the position of the particle on the ring} 
    
Conditioning again of the latest time of resetting in $[0,t[$, denoted by $t-\tau$,   we must condition on the value $\varphi$ of the angle at time $t-\tau$,
 because only trajectories for which $\varphi$ is in the safe zone survive the last resetting event:
   \begin{equation}\label{Qrtt}
   \begin{split}
    Q^{(r)}(\theta,t) =& e^{-rt} Q^{(0)}(\theta,t) + r \int_0^t d\tau e^{-r\tau}  \left(\int_\safe Q^{(r)}(\varphi, t-\tau) d\varphi\right)Q^{(0)}(\theta,\tau)\\
    =&e^{-rt} Q^{(0)}(\theta,t) + r \int_0^t d\tau e^{-r\tau}  S^{(r)}(t-\tau) Q^{(0)}(\theta,\tau), \;\;\;\;\;\;\;\;\;\;\;\theta\in]-2\pi,2\pi[.
    \end{split}
   \end{equation}
 Integrating Eq. (\ref{Qrtt}) w.r.t. $\theta$ over the safe zone yields the following renewal equation for the probability of presence in the safe zone:
 \begin{equation}\label{renSs}
    S^{(r)}(t) = e^{-rt} S^{(0)}(t) + r \int_0^t d\tau e^{-r\tau} S^{(r)}(t-\tau)  S^{(0)}(\tau).
   \end{equation}
  Taking the  Laplace transform and substituting $0$ to $s$ yields the expression of $\tilde{S}^{(r)}(0)$ in terms of quantities defined in the ordinary process:
  \begin{equation}
    \tilde{S}^{(r)}(0) = \frac{\tilde{S}^{(0)}(r)}{1 -  r \tilde{S}^{(0)}(r)}.
   \end{equation}
  Substituting into Eq. (\ref{dr0}) yields the mean first-passage time at winding number $\pm 1$ in terms of quantities defined in the ordinary process:
  \begin{equation}\label{q0rU}
   \tilde{q}^{(r)}( 0) = \tilde{q}^{(0)}( r) \left(   1 + r\frac{\tilde{S}^{(0)}(r)}{1 -  r \tilde{S}^{(0)}(r)} \right).
  \end{equation}
  It is therefore enough for our purposes to to calculate the probability density  of the walker in  the ordinary process $Q^{(0)}(\theta,t)$ for 
  all values of $\theta$ in $]-2\pi , 2\pi [$.

%

\section{Probability density of a Brownian random walker\\ without resetting}
For a Brownian random walk (in units of length and time such that the radius of the ring and the diffusion constant are both equal to $1$), the probability density $Q^{(0)}(\theta,t)$ is expressed as a  path integral:
\begin{equation}
Q^{(0)}(\theta,t) =  \int_{\Theta(0) = 0}^{\Theta(t)} D[\Theta(\tau)]\exp\left[  -\frac{1}{2}\int_0^t\left( \frac{d\Theta}{d\tau}(\tau)\right)^2 d\tau \right]
   \prod_{\tau=0}^t \mathbbm{1}\left(  \Theta(\tau) \in ]-2\pi, 2\pi[\right).
\end{equation}
where the last factor in the integrand constrains the angle $\Theta(\tau)$  to stay in the interval $]-2\pi, 2\pi[$ at all time $\tau$ in $[0,t]$. 
 We can include it in the exponential integrand as the integral over $[0,t]$ of an infinite square potential  well:
\begin{equation}
\begin{split}
Q^{(0)}(\theta,t) =&  \int_{\theta(0) = 0}^{\Theta(t)} D[\Theta(\tau)]\exp\left[   -\frac{1}{2}\int_0^t\left( \frac{d\Theta}{d\tau}(\tau)\right)^2 d\tau
  - \int_0^t V\left[  \Theta(\tau)\right] d\tau  \right],\\
 {\mathrm{where}}\;\;\;\;   V( \varphi ) =&  0 \;\;\;\;{\mathrm{if}}\;\;\varphi \in   ]-2\pi, 2\pi[, \;\;\;\;\;V( \varphi ) =+\infty\;\;\;\;\;{\mathrm{otherwise}}.
 \end{split}  
\end{equation}
Let us rewrite the probability $Q^{(0)}(\theta,t)$ using bracket notations:
\begin{equation}
\begin{split}
Q^{(0)}(\theta,t) &= \langle \theta| e^{-\hat{H}t} | 0 \rangle,\\
{\mathrm{where}}\;\;\;\;\;\;\hat{H} &= -\frac{1}{2}\frac{\partial^2}{\partial \theta^2} + V,
\end{split}
\end{equation}
A complete family of $(\psi_n)_{n\geq 1}$ normalised eigenfunctions of this Hamiltonian is given by
 \begin{equation}\label{complete}
\begin{split}
\langle  \theta | n \rangle :=& \psi_n(\theta),\;\;\;\;\;\hat{H}\psi_n = E_n \psi_n,\;\;\;\;\;\;\;(n \in {\mathbf{N}}^\ast)\\
{\mathrm{where}}\;\;\;\;\;\; \psi_n(\theta) =&\frac{1}{\sqrt{2\pi}}\sin\left(     \frac{n}{4} (\theta + 2\pi) \right),\\
{\mathrm{so}}\;\;{\mathrm{that}}\;\;\;\;\;E_n =& \frac{n^2}{32},\;\;\;\;\;\;{\mathrm{and}}\;\;\;\;\;\;\;  \sum_{n\geq 1 } | n \rangle \langle n |=1.
\end{split}
\end{equation}
 
Let us insert the identity operator into the propagator to express it in terms of the eigenfunctions:
\begin{equation}
Q^{(0)}(\theta,t) = \sum_{n\geq 1}\langle \theta| e^{-\hat{H}t}| n \rangle \langle n |  0 \rangle
       =   \sum_{n\geq 1}\langle \theta| e^{-E_n t}| n \rangle \langle n |  0 \rangle
       =  \sum_{n\geq 1} e^{-E_n t} \psi_n(\theta) \psi_n(0).
\end{equation}
We obtain the Laplace transform of the propagator in series form as
\begin{equation}\label{startFrom}
\tilde{Q}^{(0)}(\theta,s) = \sum_{n\geq 1} \frac{1}{s + E_n} \psi_n(\theta) \psi_n(0).
\end{equation}

 The Laplace transform of the survival probability reads
\begin{equation}\label{q0s}
\tilde{q}^{(0)}(s) = \int_{-2\pi}^{2\pi}\tilde{Q}^{(0)}(\theta,s) d\theta,
\end{equation}
whose explicit value  $\tilde{q}^{(0)}(s) = s^{-1}\left[  1 -  \left( \cosh\left( 2 \pi \sqrt{ 2s }\right) \right)^{-1} \right]$  was reported in \cite{KCM}.\\

Moreover, the Laplace transform of the probability of presence in the safe zone reads
\begin{equation}\label{eqWithSumsPre}
\tilde{S}^{(0)}(s) = \int_{-\pi}^{\pi}\tilde{Q}^{(0)}(\theta,s) d\theta\\.
\end{equation}
 The sums involved in the calculation of $\tilde{S}^{(0)}(s)$ can be expressed in terms of the Euler digamma function $\psi$ (the derivation is worked out in  the Appendix). 
 \begin{equation}\label{eqWithSumsRep}
\begin{split}
\tilde{S}^{(0)}(s) =&   \frac{\sqrt{2}}{8\pi s} \Psi( s),\\
{\mathrm{with}}\;\;\;\;\Psi(s) :=& \Lambda(s)+ \Lambda(s)^\ast,\\ {\mathrm{and}}\;\;\;\;\Lambda(s):=&\psi\left( \frac{1+i\sqrt{s}}{8} \right) - \psi\left( \frac{1}{8} \right)  + \psi\left( \frac{3+i\sqrt{s}}{8} \right) 
  - \psi\left( \frac{3}{8} \right)\\
   &- \psi\left( \frac{5+i\sqrt{s}}{8} \right)  +  \psi\left( \frac{5}{8} \right)
   - \psi\left( \frac{7+i\sqrt{s}}{8} \right)   + \psi\left( \frac{7}{8} \right).
   \end{split}
\end{equation}

\begin{figure}
\begin{center}
\includegraphics[width=16cm]{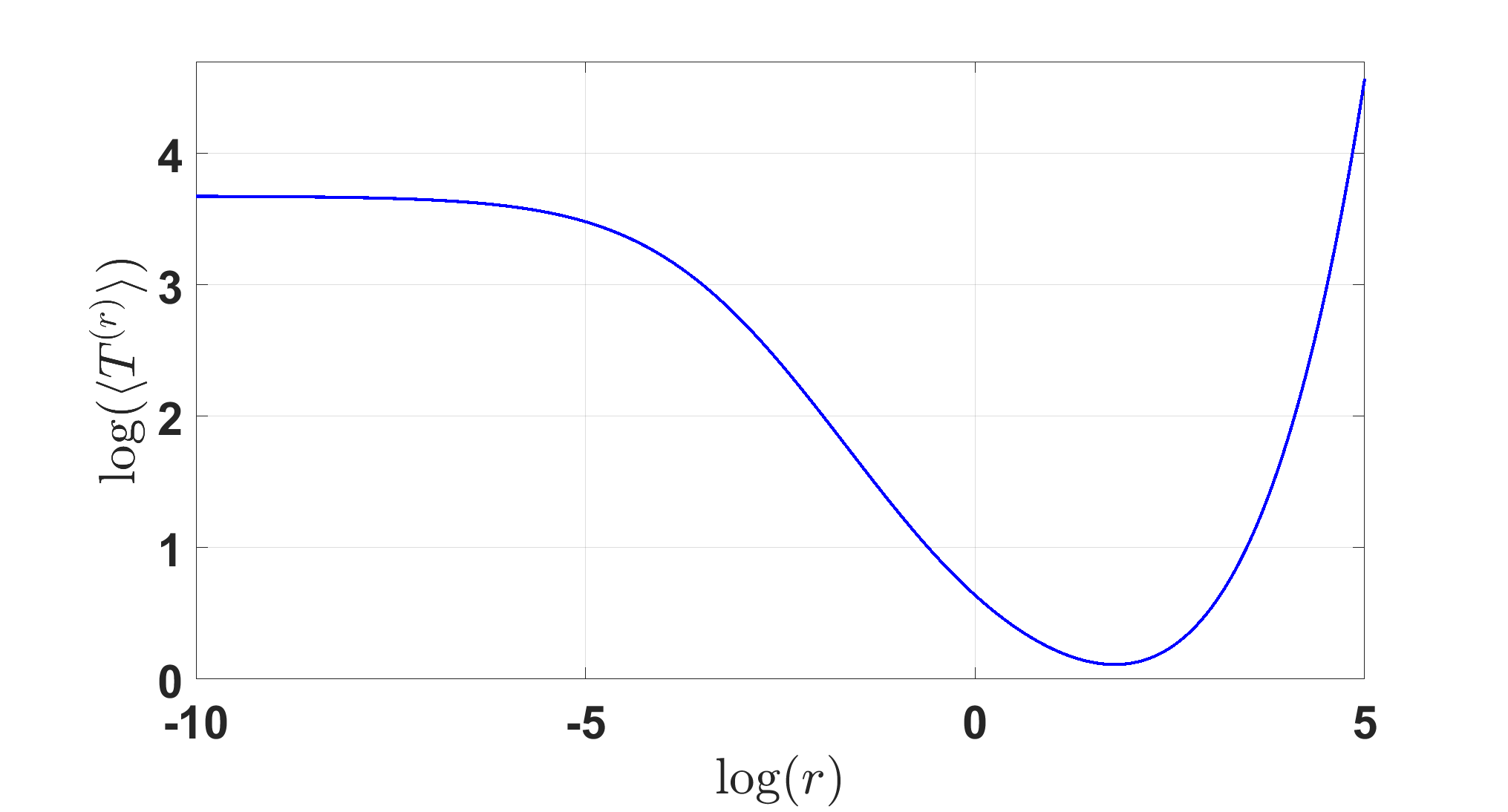} 
\caption{The mean-first passage time at winding number $\pm 1$, as a function of the resetting rate (log-log scale). Numerically the minimum
 is at $r^\ast\simeq 5.9956\dots$ (or $\log r^\ast \simeq 1.7910\dots$).}
 \label{windingOptimal}
\end{center}
\end{figure}

 From Eqs (\ref{qoi}, \ref{q0rU}), we obtain the expression of the mean first passage time  at total winding number one:
 \begin{equation}
    \langle T^{(r)} \rangle = \frac{1}{r}\left( 1 - \frac{1}{\cosh(2\pi \sqrt{2r})} \right)\left(   1- \frac{1}{4\pi\sqrt{2}}\Psi(r)\right)^{-1}.
 \end{equation}
  As $\Psi(0)=0$, this mean first-passage time has a finite limit at zero resetting rate, which reads $\tilde{q}^{(0)}(0)$, 
   the mean first-passage time in the ordinary process. At large resetting rate the mean first-passage time goes to infinity, 
    as the walker becomes less and less likely to reach the dangerous zone before being reset. The mean first-passage 
     time is plotted on Fig. \ref{windingOptimal}, where the optimal
      resetting rate is apparent. The accelerating effect of the topology in our resetting prescription leads to a lower mean first-passage time at winding number $\pm 1$ than in the ordinary process if the resetting rate is low enough.

\section{Distribution of winding numbers}

\subsection{Total winding number}
The probability density of the first-passage time  $f^{(r)}$ of the particle  at $\theta =  \pm 2\pi$ (in the process with resetting rate
 $r$) is the opposite of the time derivative of the survival probability $q^{(r)}$. Taking the Laplace transform of this relation 
  yields
  \begin{equation}
   \tilde{f}^{(r)}( s ) = - s \tilde{q}^{(r)}( s ) + 1.
  \end{equation}
  The renewal equation (Eq. (\ref{renrs}) ) satisfied by the extinction  rate  of the process, which we have only used for $s=0$ 
   so far, can be rewritten, using  Eq. (\ref{renSs}), yielding
   \begin{equation}
   \begin{split}
   \tilde{q}^{(r)}(s) &= \tilde{q}^{(0)}(r+s) \left[ 1+ r \frac{\tilde{S}^{(0)}(r+s)}{ 1 - r\tilde{S}^{(0)}(r+s) } \right],\\
    \tilde{f}^{(r)}(s) &=  1 - \frac{s  \tilde{q}^{(0)}(r+s) }{1 - r\tilde{S}^{(0)}(r+s)  }.
    \end{split}
  \end{equation}

Following the reasoning of \cite{KCM}, let us consider the event where  a Brownian particle  complete exactly $n$ turns (either clockwise or counterclockwise), under resetting at rate $r$, in the time interval $[0,t]$.  The respective durations of the turns are denoted by $\tau_1,\tau_2,\dots,\tau_n$, 
  with $\sum_{i=1}^n \tau_i < t$. The probability of this  event coincides with $n$ independent first passages at the absorbing boundaries (at times $\tau_i$, for $i\in [1..n]$),
   followed by a survival of duration $t- \sum_{i=1}^n \tau_i$.
   The joint probability density of the number of turns and their durations therefore reads
  \begin{equation}
   P( n, \{\tau_1,\dots,\tau_n\}; t) = q^{(r)}\left( t- \sum_{i=1}^n \tau_i \right)\prod_{i=1}^n  f^{(r)}( \tau_i ),\;\;\;\;\;\;n\in \mathbf{N},
  \end{equation}
 where the case $n=0$ is defined by substituting $1$ to the product and by setting the empty sum of the durations $(\tau_i)_{1\leq i \leq n}$ to zero.
  The probability law of the total number of turns at time $t$ is obtained as a convolution product by integrating over the 
   durations
   \begin{equation}
    \mathcal{P}^{(r)}(n,t) = \left(  \prod_{i=1}^n \int_0^t d\tau_i  \right) q^{(r)}\left( t- \sum_{i=1}^n \tau_i \right)\prod_{i=1}^n  f^{(r)}( \tau_i ).
   \end{equation}
 Taking the Laplace transform of this probability law maps the $(n-1)$ convolutions to ordinary products, yielding
 \begin{equation}
  \widetilde{\mathcal{P}}^{(r)}(n,s) = \tilde{q}^{(r)}( s ) \left[ \tilde{f}^{(r)}\left(s   \right)\right]^n.
 \end{equation}
 We therefore obtain the probability of the total number of turns in Laplace space in terms of quantities defined in the 
  ordinary process:
  \begin{equation}\label{prns}
  \widetilde{\mathcal{P}}^{(r)}(n,s) =  \frac{\tilde{q}^{(0)}(r+s)}{1 - r\tilde{S}^{(0)}(r+s) }\left[1 - \frac{s  \tilde{q}^{(0)}(r+s) }{1 - r\tilde{S}^{(0)}(r+s)  } \right]^n.
  \end{equation}
 The mean total number of turns follows as
 \begin{equation}
 \begin{split}
  \widetilde{\langle n \rangle} (s) & =  \sum_{n\geq 0} n  \widetilde{\mathcal{P}}^{(r)}(n,s)    \\
    & = \tilde{q}^{(r)}( s ) \frac{ \tilde{f}^{(r)}\left(s   \right)}{\left[ 1 - \tilde{f}^{(r)}\left(s   \right) \right]^2}\\
    & =  \frac{\tilde{q}^{(0)}(r+s)}{1 - r\tilde{S}^{(0)}(r+s) }\left[ 1 - \frac{s  \tilde{q}^{(0)}(r+s) }{1 - r\tilde{S}^{(0)}(r+s)  }   \right]
        \left[    \frac{1 - r\tilde{S}^{(0)}(r+s) }{s  \tilde{q}^{(0)}(r+s) }    \right]^2\\
    &= \frac{1}{s^2   \tilde{q}^{(0)}(r+s) }\left[ 1 -    r\tilde{S}^{(0)}(r+s) - s  \tilde{q}^{(0)}(r+s) \right].
  \end{split}
 \end{equation}
  The large-time behaviour of the mean number of turns can be extracted from the behaviour of the Laplace transform close to $s=0$:
 \begin{equation}
 \begin{split}
  \widetilde{\langle n \rangle} (s) & =  \frac{1}{s^2   \tilde{q}^{(0)}(r+s) }\left[ 1 -    r\tilde{S}^{(0)}(r+s) - s  \tilde{q}^{(0)}(r+s) \right]\\       
  &= \frac{1}{s^2   \tilde{q}^{(0)}(r) }\left[   1 - r \tilde{S}^{(0)}(r) 
    -s \left(  \tilde{q}^{(0)}(r)  + r (\tilde{S}^{(0)})'(r) + \frac{(\tilde{q}^{(0)})'(r) }{\tilde{q}^{(0)}(r) } \right)    + o( s )\right],
  \end{split}
 \end{equation}
 \begin{equation}
 \begin{split}
\langle n \rangle  (t) &\underset{t\to\infty}{=}   W_r t + w_r + o(1)\\
W_r &=  \frac{1 - r \tilde{S}^{(0)}(r)   }{\tilde{q}^{(0)}(r)},\;\;\;\;\;\;\;\;\;\;\;\;\;\;\;\;\;w_r = -1 - r \frac{(\tilde{S}^{(0)})'(r)}{  \tilde{q}^{(0)}(r)} - \frac{(\tilde{q}^{(0)})'(r) }{\left( \tilde{q}^{(0)}(r) \right)^2}.
\end{split}
 \end{equation}

\subsection{Net winding number}
 
If the walker has completed $n_+$ counterclockwise turns and $n_-$ clockwise turns (in any order), 
 the total winding number is $n:=n_+ + n_-$, and the net winding number is $k:= n_+ - n_-$. The probability of having 
  net winding number $k$ and total winding number $n$ at time $t$ (denoted by $P(n,k,t)$) is the sum of the probabilities of any sequence 
   of turns of independent signs, exactly $n_+ =  ( n+ k )/2$ of which are counterclockwise. Each of the signs in the sequence has probability $1/2$
    to occur, hence
    \begin{equation}
     \mathcal{P}^{(r)}( n,k,t) = \binom{n}{\frac{n+k}{2}} \left(\frac{1}{2}\right)^n \mathcal{P}^{(r)}(n,t), \;\;\;\;\;\;\;\;(n\in \mathbf{N}, k \in [-n..n]).
    \end{equation} 
    The marginal probability $P(k,t)$  of the net winding number $k$ at time $t$  is even in $k$, hence:
 \begin{equation}
 \mathcal{P}^{(r)}( k, t ) = \sum_{m \geq 0}\mathcal{P}^{(r)}( 2m + |k|, k, t )= \sum_{m \geq 0} \binom{ 2m + |k|}{k} \left(\frac{1}{2}\right)^{2m + |k|} \mathcal{P}^{(r)}(n,t),\;\;\;\;\;\;\;\;\;\;\;\;\;\;\; (k\in \mathbf{Z}).
\end{equation} 
 In Laplace space, Eq. (\ref{prns}) yields 
 \begin{equation}
 \begin{split}
 \widetilde{\mathcal{P}}^{(r)}( k, s ) =& \sum_{m \geq 0} \binom{ 2m + |k|}{k} \left(\frac{1}{2}\right)^{2m + |k|}  \frac{\tilde{q}^{(0)}(r+s)}{1 - r\tilde{S}^{(0)}(r+s) }\left[1 - \frac{s  \tilde{q}^{(0)}(r+s) }{1 - r\tilde{S}^{(0)}(r+s)  } \right]^{ 2m + |k|}\\ 
  =& \frac{\tilde{q}^{(0)}(r+s)}{1 - r\tilde{S}^{(0)}(r+s) }\left[\frac{1}{2} - \frac{s  \tilde{q}^{(0)}(r+s) }{2[ 1 - r\tilde{S}^{(0)}(r+s) ]  } \right]^{ |k|}\\ 
        &\times \sum_{m \geq 0} \binom{ 2m + |k|}{k} \left[ \frac{1}{4}\left( 1 - \frac{s  \tilde{q}^{(0)}(r+s) }{ 1 - r\tilde{S}^{(0)}(r+s)}  \right)^2 \right]^m \\
 =&   \frac{\tilde{q}^{(0)}(r+s)}{1 - r\tilde{S}^{(0)}(r+s) }z^{\frac{|k|}{2}}\frac{1}{\sqrt{1-4z}}\left( \frac{1 - \sqrt{1-4z}}{2z}   \right)^{|k|}\;\;\;\;\;\;\;\;\;\;\;\;\;\;\; (k\in \mathbf{Z}),
 \end{split}
\end{equation} 
 with the notation
 \begin{equation}
  z(s) = \frac{1}{4}\left( 1 - \frac{s  \tilde{q}^{(0)}(r+s) }{ 1 - r\tilde{S}^{(0)}(r+s)}  \right)^2.
 \end{equation}
 We used the identity borrowed from \cite{KCM}, where it was 
  used in the same context (for $|z|<1/4$):
  \begin{equation}
      \sum_{m \geq 0} \binom{ 2m + |k|}{k} z^m = \frac{1}{\sqrt{1-4z}}\left( \frac{1 - \sqrt{1-4z}}{2z}   \right)^{|k|}.
  \end{equation}

 By symmetry $\langle k \rangle$ is identically zero. Hence the variance of the net winding number
  is obtained as
  \begin{equation}
 \begin{split}
 \widetilde{\langle k^2 \rangle}(s) =&   \frac{2\tilde{q}^{(0)}(r+s)}{1 - r\tilde{S}^{(0)}(r+s) }    \frac{1}{\sqrt{1-4z(s)}}\sum_{k\geq 0} k^2\left(   \frac{1 - \sqrt{1-4z(s)}}{2\sqrt{z(s)}}     \right)^k  \\
 =&  \frac{2\tilde{q}^{(0)}(r+s)}{1 - r\tilde{S}^{(0)}(r+s) }    \frac{1}{\sqrt{1-4z(s)}}\frac{Z(s)( 1+Z(s))}{(1-Z(s))^3},\\
 {\mathrm{with}}&\;\;\;\;\;\;\;\; Z:= \frac{1 - \sqrt{1-4z}}{2\sqrt{z}} = \frac{1- \sqrt{sK(s)(2-sK(s))}}{1 - s K(s) },\;\;\;\;\;\;\;
  K(s):=\frac{\tilde{q}^{(0)}(r+s)}{1 - r\tilde{S}^{(0)}(r+s) }.
 \end{split}
\end{equation} 
Substituting and factorising the leading contribution at small $s$ yields 
  \begin{equation}
 \begin{split}
 \widetilde{\langle k^2 \rangle}(s)   
 =& \frac{1}{s^2K(s)}\frac{ (1-sK(s))   \left[1-sK(s)/2 - \sqrt{\frac{sK(s)}{2}\left( 1 - \frac{sK(s)}{2}\right)}\right]     [1-   \sqrt{sK(s)( 2 - sK(s))}]   }{\left( \sqrt{1-\frac{sK(s)}{2}}  - \sqrt{\frac{sK(s)}{2}}\right)^3\sqrt{1 - \frac{sK(s)}{2}}}\\
 &\underset{s\to 0}{\sim}\frac{1}{s^2}\frac{1 - r \tilde{S}^{(0)}(r)   }{\tilde{q}^{(0)}(r)}.
 \end{split}
\end{equation} 
Inverting the leading term of the Laplace transform close to $s=0$ yields the large-time equivalent
 \begin{equation}
 {\langle k^2 \rangle}(t) \underset{t\to \infty}{\sim} W_r t.
 \end{equation}
For any underlying random walk in the ordinary process (not just the  diffusive random walk), the large-time rate of increase of the variance of the net winding number (or the mean total winding number), is given by the inverse of the mean first-completion time of a turn:
\begin{equation}
 W_r =  \frac{1}{\langle  T^{(r)}\rangle}.
\end{equation}
 In the case of a diffusive random walk, the coefficient $W_r$ has a finite limit when the resetting rate $r$ goes to zero:
 \begin{equation}
  \underset{r\to 0+}{\lim} W_r = \frac{1}{4\pi^2},
 \end{equation}
  which corresponds to the large-time equivalents of $\langle n  \rangle(t)$ and $\langle k^2 \rangle(t)$ reported in \cite{KCM}.

 \section{Discussion}

 We have obtained a system of renewal equations for the probability density and survival rate of a random walker on a ring 
  with absorbing boundary conditions at winding number $\pm1$,  subjected to Poisson-distributed resetting to its initial position. 
   The process is accelerated if the walker is reset when its polar angle is within $\pi$ radians
    of the completion of a turn. We have expressed the mean completion time of a turn in terms of the probability density of the random walker in the 
     ordinary process (without resetting). This relation holds for any process underlying the random walk. In the Brownian case 
      we have obtained the mean completion time of a turn, and put in a closed form involving the digamma function. The inverse of this time 
       is the rate of increase of the mean total winding number (and of the variance of the net winding number) at long time. \\

 We have chosen a system of units in which both the radius $R$  of the ring and the diffusion constant $D$ are set to $1$.
  For dimensional reasons, we can restore these parameters and find the optimal value of the resetting rate 
    close to $5.9956 R^{-2}D$. In the limit of a large radius, this optimal value of the resetting rate becomes small, because the topological  benefits of acceleration 
    are felt only when at least one half of a turn  has been completed during a renewal period (which becomes  less probable at fixed resetting rate 
     if the radius becomes larger). \\

 There are two major differences with the case of a diffusive random walker on a non-compact one-dimensional space: 
  the mean first-passage time is finite in the ordinary case on a ring, and the system under resetting does not reach a steady state.      
      The acceleration of the process occurs for topological reasons: in our resetting prescription, the walker takes the shortest possible path
       to the physical starting point of the process (whose polar coordinate can take any value among the integer multiples of $2\pi$). Possible generalisations of the present work include absorbing boundary conditions at fixed winding numbers $-w_-$ and $w_+$, with $w_\pm>1$ (the dangerous zone still being within $\pi$ radians of each boundary). 
       Moreover, it would be interesting to 
        estimate the behaviour of the winding number of random walks in the plane with a singularity at the origin,  subjecting to resetting
         the models of polymers studied in \cite{edwards1967statistical,edwards1968statistical,inomata1978path,bernido1981path}. \\

       Developments 
       on the experimental realisation of stochastic resetting (using optical techniques) have been reported in \cite{experimentalResetting,besga2020optimal,faisant2021optimal}, with extension to two dimensions and investigation of periodic resetting protocols.
      An experimental realisation of our resetting prescription for a random walker on a ring could be provided by a one-armed mechanical clock, suspended vertically. The random walker would sit at the tip of the arm, the process would start in a configuration where the arm indicates 6 o'clock. The arm would turn by independent random amounts generated by the random walk and transmitted through a rod at the center of the clock. At resetting times the connection between the arm and the rod  would become loose, and the arm would go back to its initial position under the influence of gravity (this move of the arm is considered to be instantaneous in our model). It would be natural to compare this mechanism to the realisation of resetting by confining potentials, as introduced in \cite{gupta2021resetting}. It would  be interesting to generalise the model to random walks on higher-dimensional non-simply-connected spaces.\\


 \section*{Appendix}
 
Let us work out the  expression of the probability density of the position of the  diffusive random walker without resetting,  starting from  the Laplace transform of the  Eq. (\ref{startFrom}) and the definitions 
  of Eq. (\ref{complete}):
\begin{equation}\label{eqWithSums}
\begin{split}
\tilde{Q}^{(0)}(\theta,s) =&  \sum_{n\geq 1} \frac{1}{s + E_n} \psi_n(\theta) \psi_n(0)\\
  =&   \frac{1}{2\pi}  \sum_{n\geq 1} \frac{1}{s + \frac{n^2}{32}}\sin\left(   \frac{n\theta}{4} +  \frac{n\pi}{2} \right)
   \sin\left(    \frac{n\pi }{2} \right)\\
  =&   \frac{1}{4\pi}  \sum_{n\geq 1} \frac{1}{s + \frac{n^2}{32}}  \left[  \cos\left(  \frac{n\theta}{4 } \right) 
  -\cos\left(  \frac{n\theta}{4 } +  n\pi  \right)  \right]\\\ 
   =& \frac{16}{\pi}\sum_{m\geq 0} \frac{1}{32 s + (2m+1)^2}   \cos\left(  \frac{(2m+1)\theta}{4} \right).\\
\end{split}
\end{equation}
 
We find the Laplace transform of the survival probability as
\begin{equation}\label{q0s}
\begin{split}
\tilde{q}^{(0)}(s) &= \int_{-2\pi}^{2\pi}\tilde{Q}^{(0)}(\theta,s) d\theta\\
    &=  \frac{1}{2\pi}\sum_{m\geq 0} \frac{1}{s + \frac{(2m+1)^2}{32}}\frac{8}{2m + 1} \sin\left(\frac{(2m+1)\pi}{2} \right)\\
    &=  \frac{4}{\pi}\sum_{m\geq 0} \frac{32}{32 s + (2m+1)^2}\frac{(-1)^m}{2m + 1},\\
   \end{split}
\end{equation}
which coincides wth the value $\frac{1}{s} \left[  1 - \frac{1}{\cosh\left( 2 \pi \sqrt{ 2s } \right)}  \right]$ obtained in \cite{KCM}.\\

 The Laplace transform of the probability of presence in the safe zone reads
\begin{equation}\label{eqWithSumsPre}
\begin{split}
\tilde{S}^{(0)}(s) = \int_{-\pi}^{\pi}\tilde{Q}^{(0)}&(\theta,s) d\theta\\
   =  \frac{1}{2\pi}&\sum_{m\geq 0} \frac{1}{s + \frac{(2m+1)^2}{32}}\frac{8}{2m + 1} \sin\left(\frac{(2m+1)\pi}{4} \right)\\
   = \frac{4}{\pi\sqrt{2}}  &   \sum_{m = 0,1 + 4k, k\geq 0} \frac{32}{32s + (2m+1)^2}\frac{1}{2m + 1}           
       -   \frac{4}{\pi\sqrt{2}} \sum_{m = 2, 3 + 4k, k\geq 0} \frac{32}{32 s + (2m+1)^2}\frac{1}{2m + 1}  \\
   = \frac{64\sqrt{2}}{\pi} &\sum_{k\geq 0}\left[ \frac{1}{[32 s + (8k+1)^2](8k + 1)}  + \frac{1}{[32 s + (8k+3)^2](8k + 3)}      \right.\\ 
        &\;\;\;\;\; \left.  -   \frac{1}{[32 s + (8k+5)^2](8k + 5)}  - \frac{1}{[32 s + (8k+7)^2](8k + 7)}   \right].\\
   \end{split}
\end{equation}

 To work out the sums involved in Eq. (\ref{eqWithSums}), let us start from the identity (labelled 6.3.16 in  \cite{abramowitz1988handbook})
\begin{equation}\label{identity}
\psi( z ) = -\gamma + \sum_{n\geq 0 } \left( \frac{1}{ n + 1} - \frac{1}{ n + z } \right),
\end{equation}
where $\psi = \Gamma'/\Gamma$ denotes the Euler digamma function.
 Consider the following combination:
 \begin{equation}
  C_k(u):= \psi\left( \frac{k+i\sqrt{u}}{8} \right) + \psi\left( \frac{k-i\sqrt{u}}{8} \right) -2 \psi\left( \frac{k}{8} \right).
 \end{equation}
  where for our purposes $u$ will be a positive number and $k$ will be an integer.
 \begin{equation}
 \begin{split}
  C_k(u) =& \sum_{n\geq 0}\left[ \frac{8}{8n + k + i\sqrt{u}} + \frac{8}{8n + k - i\sqrt{u}}  - \frac{16}{8n + k}\right]\\
  =& \sum_{n\geq 0}\frac{16u}{(u + (8n + k )^2)(8n + k)}.
  \end{split}
 \end{equation}
 
%
Eq. (\ref{eqWithSumsPre}) therefore reads 
\begin{equation}\label{refSum}
\begin{split}
 \tilde{S}^{(0)}(s) =&\frac{\sqrt{2}}{8\pi s}\left[ C_1(32 s) +   C_3(32 s) -  C_5(32 s) -  C_7(32 s) \right]\\ 
  =&\frac{\sqrt{2}}{8\pi s}\left[   \psi\left(  \frac{k+i\sqrt{u}}{8} \right) + \psi\left( \frac{k-i\sqrt{u}}{8} \right) -2 \psi\left( \frac{1}{8}\right) \right.\\
             &  +   \psi\left(     \frac{3+i\sqrt{u}}{8} \right) + \psi\left( \frac{3-i\sqrt{u}}{8} \right) -2 \psi\left( \frac{3}{8}\right)                             \\
             &  -      \psi\left(      \frac{5+i\sqrt{u}}{8} \right) - \psi\left( \frac{5-i\sqrt{u}}{8} \right) +2 \psi\left( \frac{5}{8}\right)                           \\
              &  \left.       - \psi\left(  \frac{7+i\sqrt{u}}{8} \right) - \psi\left( \frac{7-i\sqrt{u}}{8} \right) +2 \psi\left( \frac{7}{8}\right)          \right].
\end{split}   
\end{equation}
 Using the identity $\psi( z) = \overline{\psi( z )}$ (which is manifest from Eq. (\ref{identity})) yields the expression reported in Eq. (\ref{eqWithSumsRep}).

\bibliography{bibRefsNew} 

\begin{thebibliography}{10}

\bibitem{evans2011diffusion}
M.~R. Evans and S.~N. Majumdar, ``Diffusion with stochastic resetting,'' {\em
  Physical review letters}, vol.~106, no.~16, p.~160601, 2011.

\bibitem{evans2011optimal}
M.~R. Evans and S.~N. Majumdar, ``Diffusion with optimal resetting,'' {\em
  Journal of Physics A: Mathematical and Theoretical}, vol.~44, no.~43,
  p.~435001, 2011.

\bibitem{evans2018run}
M.~R. Evans and S.~N. Majumdar, ``Run and tumble particle under resetting: a
  renewal approach,'' {\em Journal of Physics A: Mathematical and Theoretical},
  vol.~51, no.~47, p.~475003, 2018.

\bibitem{refractory}
M.~R. Evans and S.~N. Majumdar, ``Effects of refractory period on stochastic
  resetting,'' {\em Journal of Physics A: Mathematical and Theoretical},
  vol.~52, no.~1, p.~01LT01, 2018.

\bibitem{magoni2020ising}
M.~Magoni, S.~N. Majumdar, and G.~Schehr, ``Ising model with stochastic
  resetting,'' {\em Physical Review Research}, vol.~2, no.~3, p.~033182, 2020.

\bibitem{grange2020entropy}
P.~Grange, ``Entropy barriers and accelerated relaxation under resetting,''
  {\em Journal of Physics A: Mathematical and Theoretical}, 2020.

\bibitem{mercado2018lotka}
G.~Mercado-V{\'a}squez and D.~Boyer, ``Lotka--{V}olterra systems with
  stochastic resetting,'' {\em Journal of Physics A: Mathematical and
  Theoretical}, vol.~51, no.~40, p.~405601, 2018.

\bibitem{toledo2019predator}
J.~Q. Toledo-Marin, D.~Boyer, and F.~J. Sevilla, ``Predator-prey dynamics:
  Chasing by stochastic resetting,'' {\em arXiv preprint arXiv:1912.02141},
  2019.

\bibitem{ZRPSS}
P.~Grange, ``Steady states in a non-conserving zero-range process with
  extensive rates as a model for the balance of selection and mutation,'' {\em
  Journal of Physics A: Mathematical and Theoretical}, vol.~52, no.~36,
  p.~365601, 2019.

\bibitem{ZRPResetting}
P.~Grange, ``Non-conserving zero-range processes with extensive rates under
  resetting,'' {\em Journal of Physics Communications}, vol.~4, no.~4,
  p.~045006, 2020.

\bibitem{durang2014statistical}
X.~Durang, M.~Henkel, and H.~Park, ``The statistical mechanics of the
  coagulation--diffusion process with a stochastic reset,'' {\em Journal of
  Physics A: Mathematical and Theoretical}, vol.~47, no.~4, p.~045002, 2014.

\bibitem{grange2021aggregation}
P.~Grange, ``Aggregation with constant kernel under stochastic resetting,''
  {\em Journal of Physics A: Mathematical and Theoretical}, 2021.

\bibitem{lapeyre2019stochastic}
G.~J. Lapeyre~Jr and M.~Dentz, ``Stochastic processes under reset,'' {\em arXiv
  preprint arXiv:1903.08055}, 2019.

\bibitem{gupta2018stochastic}
D.~Gupta, ``Stochastic resetting in underdamped {B}rownian motion,'' {\em
  Journal of Statistical Mechanics: Theory and Experiment}, vol.~2019, no.~3,
  p.~033212, 2019.

\bibitem{basu2019long}
U.~Basu, S.~N. Majumdar, A.~Rosso, and G.~Schehr, ``Long time position
  distribution of an active brownian particle in two dimensions,'' {\em arXiv
  preprint arXiv:1908.10624}, 2019.

\bibitem{basu2019symmetric}
U.~Basu, A.~Kundu, and A.~Pal, ``Symmetric exclusion process under stochastic
  resetting,'' {\em Physical Review E}, vol.~100, no.~3, p.~032136, 2019.

\bibitem{pal2021inspection}
A.~Pal, S.~Kostinski, and S.~Reuveni, ``The inspection paradox in stochastic
  resetting,'' {\em arXiv preprint arXiv:2108.07018}, 2021.

\bibitem{bressloff2020directed}
P.~C. Bressloff, ``Directed intermittent search with stochastic resetting,''
  {\em Journal of Physics A: Mathematical and Theoretical}, vol.~53, no.~10,
  p.~105001, 2020.

\bibitem{bressloff2020modeling}
P.~C. Bressloff, ``Modeling active cellular transport as a directed search
  process with stochastic resetting and delays,'' {\em Journal of Physics A:
  Mathematical and Theoretical}, vol.~53, no.~35, p.~355001, 2020.

\bibitem{singh2020random}
P.~Singh, ``Random acceleration process under stochastic resetting,'' {\em
  Journal of Physics A: Mathematical and Theoretical}, vol.~53, no.~40,
  p.~405005, 2020.

\bibitem{majumdar2021mean}
S.~N. Majumdar, F.~Mori, H.~Schawe, and G.~Schehr, ``Mean perimeter and area of
  the convex hull of a planar {B}rownian motion in the presence of resetting,''
  {\em Physical Review E}, vol.~103, no.~2, p.~022135, 2021.

\bibitem{topical}
M.~R. Evans, S.~N. Majumdar, and G.~Schehr, ``Stochastic resetting and
  applications,'' {\em arXiv preprint arXiv:1910.07993}, 2019.

\bibitem{KCM}
A.~Kundu, A.~Comtet, and S.~N. Majumdar, ``Winding statistics of a {B}rownian
  particle on a ring,'' {\em Journal of Physics A: Mathematical and
  Theoretical}, vol.~47, no.~38, p.~385001, 2014.

\bibitem{calvert2021searching}
G.~R. Calvert and M.~R. Evans, ``Searching for clusters of targets under
  stochastic resetting,'' {\em The European Physical Journal B}, vol.~94,
  no.~11, pp.~1--9, 2021.

\bibitem{edwards1967statistical}
S.~F. Edwards, ``Statistical mechanics with topological constraints: I,'' {\em
  Proceedings of the Physical Society (1958-1967)}, vol.~91, no.~3, p.~513,
  1967.

\bibitem{edwards1968statistical}
S.~Edwards, ``Statistical mechanics with topological constraints: Ii,'' {\em
  Journal of Physics A: General Physics}, vol.~1, no.~1, p.~15, 1968.

\bibitem{inomata1978path}
A.~Inomata and V.~A. Singh, ``Path integrals with a periodic constraint:
  Entangled strings,'' {\em Journal of Mathematical Physics}, vol.~19, no.~11,
  pp.~2318--2323, 1978.

\bibitem{bernido1981path}
C.~C. Bernido and A.~Inomata, ``Path integrals with a periodic constraint: The
  {A}haronov--{B}ohm effect,'' {\em Journal of Mathematical physics}, vol.~22,
  no.~4, pp.~715--718, 1981.

\bibitem{experimentalResetting}
O.~Tal-Friedman, A.~Pal, A.~Sekhon, S.~Reuveni, and Y.~Roichman, ``Experimental
  realization of diffusion with stochastic resetting,'' {\em arXiv preprint
  arXiv:2003.03096}, 2020.

\bibitem{besga2020optimal}
B.~Besga, A.~Bovon, A.~Petrosyan, S.~N. Majumdar, and S.~Ciliberto, ``Optimal
  mean first-passage time for a {B}rownian searcher subjected to resetting:
  experimental and theoretical results,'' {\em Physical Review Research},
  vol.~2, no.~3, p.~032029, 2020.

\bibitem{faisant2021optimal}
F.~Faisant, B.~Besga, A.~Petrosyan, S.~Ciliberto, and S.~N. Majumdar, ``Optimal
  mean first-passage time of a {B}rownian searcher with resetting in one and
  two dimensions: Experiments, theory and numerical tests,'' {\em arXiv
  preprint arXiv:2106.09113}, 2021.

\bibitem{gupta2021resetting}
D.~Gupta, A.~Pal, and A.~Kundu, ``Resetting with stochastic return through
  linear confining potential,'' {\em Journal of Statistical Mechanics: Theory
  and Experiment}, vol.~2021, no.~4, p.~043202, 2021.

\bibitem{abramowitz1988handbook}
M.~Abramowitz, I.~A. Stegun, and R.~H. Romer, ``Handbook of mathematical
  functions with formulas, graphs, and mathematical tables,'' 1988.

\end{thebibliography}
\bibliographystyle{ieeetr}

\end{document}